\documentclass[aps,prl,twocolumn,groupedaddress,amsmath,amssymb]{revtex4}

\usepackage{graphicx}
\usepackage{SIunits}

\usepackage[letterpaper,pdftex]{hyperref}
\hypersetup{
	colorlinks=true,
	linkcolor=black,
	citecolor=black,
        urlcolor=blue
}

\begin{document}


\title{Reversible switching of surface texture by hydrogen intercalation}


\author{Thomas Brugger}
\affiliation{Physik-Institut, Universit\"{a}t Z\"{u}rich, Winterthurerstrasse 190, CH-8057 Z\"{u}rich, Switzerland}
\author{Haifeng Ma}
\affiliation{Physik-Institut, Universit\"{a}t Z\"{u}rich, Winterthurerstrasse 190, CH-8057 Z\"{u}rich, Switzerland}
\author{Marcella Iannuzzi}
\affiliation{Physikalisch Chemisches Institut, Universit\"{a}t Z\"{u}rich, Winterthurerstrasse 190, CH-8057 Z\"{u}rich, Switzerland}
\author{Simon Berner}
\affiliation{Physik-Institut, Universit\"{a}t Z\"{u}rich, Winterthurerstrasse 190, CH-8057 Z\"{u}rich, Switzerland}
\author{Adolf Winkler}
\affiliation{Institute of Solid State Physics, Graz University of Technology, Petersgasse 16, A-8010 Graz, Austria}
\author{J\"{u}rg Hutter}
\affiliation{Physikalisch Chemisches Institut, Universit\"{a}t Z\"{u}rich, Winterthurerstrasse 190, CH-8057 Z\"{u}rich, Switzerland}
\author{ J\"{u}rg Osterwalder}
\affiliation{Physik-Institut, Universit\"{a}t Z\"{u}rich, Winterthurerstrasse 190, CH-8057 Z\"{u}rich, Switzerland}
\author{Thomas Greber}
\email{greber@physik.uzh.ch}
\affiliation{Physik-Institut, Universit\"{a}t Z\"{u}rich, Winterthurerstrasse 190, CH-8057 Z\"{u}rich, Switzerland}


\date{\today}

\begin{abstract}
The interaction of atomic hydrogen with a single layer of hexagonal boron nitride on rhodium leads to a removal of the $h$-BN surface corrugation. The process is reversible as the hydrogen may be expelled by annealing to about $\unit{500}{\kelvin}$ whereupon the texture on the nanometer scale is restored. This effect is traced back to hydrogen intercalation. It is expected to have implications for applications, like the storage of hydrogen, the peeling of sp$^2$-hybridized layers from solid substrates or the control of the wetting angle, to name a few.
\end{abstract}


\maketitle


\section{Introduction}
Intercalation -- that is the reversible embedding of atomic or molecular species into a layered material -- is a key concept for materials functionalization. Graphite is the prototype intercalation material \cite{dre81} in which the relatively weak bonding between sp$^2$-hybridized carbon sheets allows the packing of molecular species between subsequent layers. On transition metals, single sheets of carbon (graphene) and boron nitride ($h$-BN) form highly ordered nanostructures  \cite{lan92,osh97,cor04,pan07}. The lateral periodicity of these superstructures is determined by the lattice mismatch between the sp$^2$ layers and the substrate. For 4d and 5d transition metals the superlattice constant is about 10 times that of the free standing sp$^2$ layer \cite{lan92,osh97,cor04,pan07}. The deformation of the surfaces in the vertical direction -- the {\it {corrugation}} -- can reach values of $\unit{0.1}{\nano\meter}$ and is controlled by the bonding of B, C and N to the transition metals, where the lattice mismatch {\it {and}} the site dependent bonding impose a dislocation network with mainly out-of-plane strain in the sp$^2$ layer. The corrugation is the essential property that determines the texture and functionality of these superstructures with a variety of new phenomena. These are for example the formation of dipole rings which are in-plane lateral electric fields \cite{dil08,bru09} that constitute traps for single molecules or clusters at room temperature \cite{ber07,zha08}.  Here it is shown that the surface texture of $h$-BN/Rh(111) may be reversibly switched from corrugated to flat by hydrogen intercalation and back to corrugated by hydrogen removal.

\section{Results}
\begin{figure}
\includegraphics[width=\columnwidth]{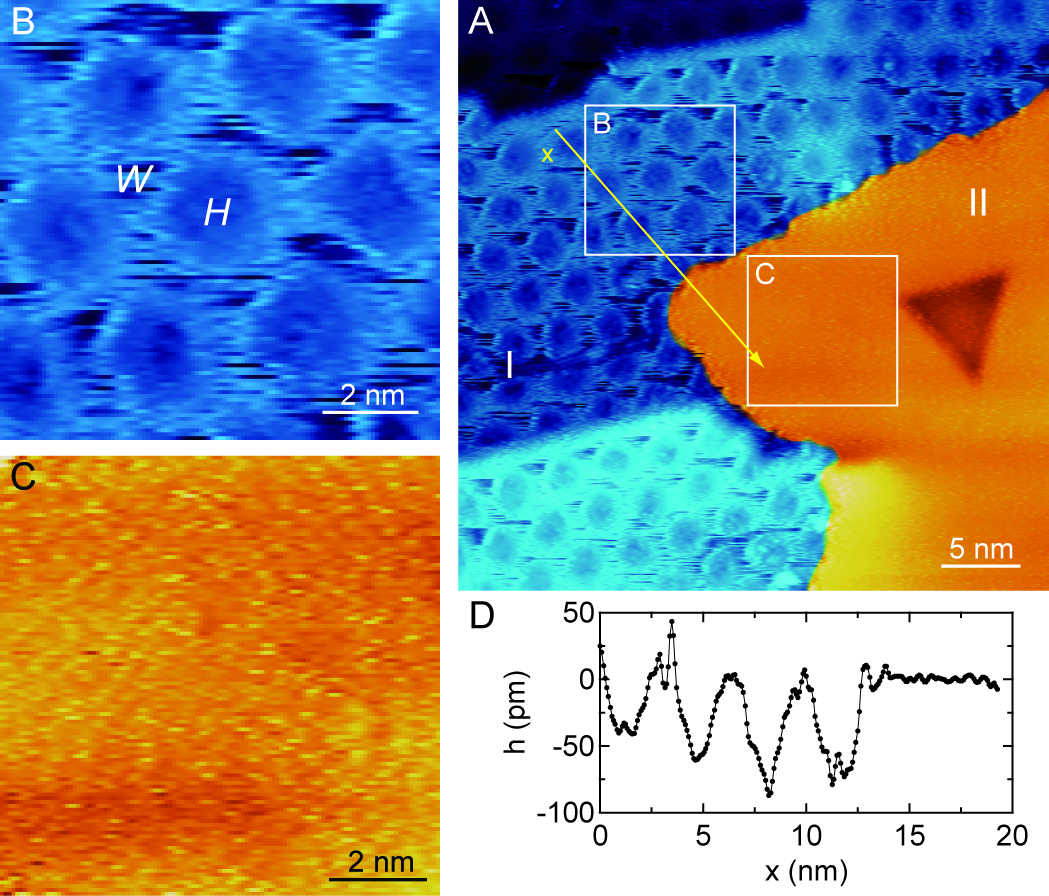}
\caption{\label{F:HhBNRh_STM}Topographic STM data of the $h$-BN/Rh(111) after exposure to atomic hydrogen. A) Large scale image showing the coexistence of pristine corrugated and hydrogen intercalated flat $h$-BN/Rh(111) (regions I and II respectively). B) Zoom into the pristine $h$-BN/Rh(111) area showing the wire ($W$) and the hole ($H$) regions. C) Zoom into the flat $h$-BN/Rh(111) area. D) Line profile along the yellow line in a) showing the transition from corrugated (I) to flat (II) $h$-BN/Rh(111).}
\end{figure}
The corrugation of the $h$-BN/Rh(111) nanomesh superstructure can directly be observed using scanning tunneling microscopy (STM). It has a peculiar electronic structure which is reflected in the core level binding energies of the elements that are sensitive to the position of the atoms  \cite{pre07b}. A splitting of the N 1s core level into two components as observed with x-ray photoelectron spectroscopy (XPS) is ascribed to the $h$-BN layer corrugation \cite{pre07b} and a separate set of sp$^2$-derived bands shows up in ultraviolet photoelectron spectroscopy (UPS) \cite{cor04}. 

The topographic STM data presented in Figure \ref{F:HhBNRh_STM} show that after exposure to atomic hydrogen (H) a new phase of single layer $h$-BN without long-range periodic corrugation (II) can coexist with the unaltered nanomesh (I). Phase II is basically flat whereas the pristine nanomesh (I) exhibits a corrugated hexagonal surface texture with a periodicity of $\unit{3.2}{\nano\meter}$ \footnote{A third phase which has been observed in further STM images (not shown) consists of a crumpled $h$-BN layer in which the original nanomesh is still visible at a few places. This phase is the intermediate level where insufficient H intercalation led to a nanomesh corrugation which is not yet fully switched off.}. The dark triangle in region II of Figure \ref{F:HhBNRh_STM} A) is a dislocation within the Rh(111) substrate as it is often found and was used as a landmark for orientation. The zoom-in images in Figure \ref{F:HhBNRh_STM} display a section of the same area taken from regions with and without corrugation. In the textured zoom-in six nanomesh unit cells are seen, where the two topographic elements, the `holes' ($H$) and the `wires' ($W$) become visible (Fig. \ref{F:HhBNRh_STM} B). The zoom into the untextured region is basically flat (Fig. \ref{F:HhBNRh_STM} C). The height profile in Figure \ref{F:HhBNRh_STM} D) shows the transition from the intact, corrugated nanomesh region into the flat region. The surface in the flat region levels on the height of the wires that is the loosely bound regions in the nanomesh.

\begin{figure}
\includegraphics[width=\columnwidth]{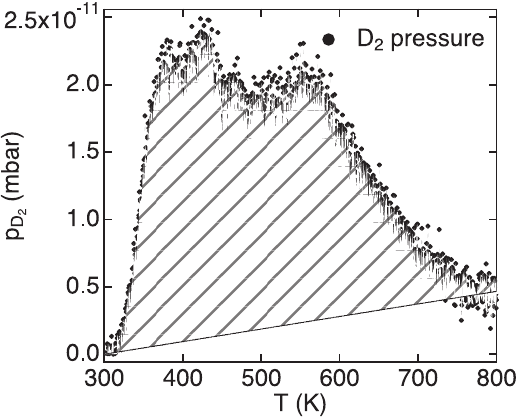}
\caption{\label{F:HhBNRh_TDS}Deuterium TDS data from $h$-BN/Rh(111) intercalated with D (filled black circles, $\beta=\unit{0.2}{\kelvin\rp\second}$). The hatched area is proportional to the amount of D intercalated between the $h$-BN and the topmost Rh layer.}
\end{figure}
In order to determine the amount of H that is incorporated into $h$-BN/Rh(111) in the switched off (flat) state we performed thermal desorption spectroscopy (TDS) measurements during a switch on (roughening) process of the $h$-BN/Rh(111) nano-texture by annealing. We found no isotope effect between hydrogen and deuterium (D), though it is advantageous to use D since the deuterium background signal in TDS is much smaller than that of hydrogen. Also the use of isotopes allows  to show that the effect is caused by atomic H (D), and  at the investigated conditions not by its molecular forms  H$_2$ (D$_2$). Figure \ref{F:HhBNRh_TDS} shows the TDS curve of molecular deuterium (D$_2$, atomic mass 4, heating rate $\beta=\unit{0.2}{\kelvin\rp\second}$) from a sample that had been exposed to D to switch off the surface texture and subsequently flushed with H$_2$ in order to supplant D on the sample holder with an isotope exchange reaction. Simultaneously the TDS signal of HD (atomic mass 3) has been recorded. From the area under both curves ($\unit{3}{amu}$ and $\unit{4}{amu}$) and the pumping speed the amount $n_\text{D}$ of desorbed D can be inferred. We find $n_\text{D}\approx(\unit{4\pm0.5)\cdot\power{10}{14}}{\centi\rp\meter}$ D atoms to be incorporated in the untextured state. This value of $n_\text{D}$ corresponds to about $\unit{0.25}{ML}$ ($\unit{1}{ML}\equiv$ one atom per Rh top layer atom) but has to be regarded as a lower limit for the incorporated D. Remarkably, $n_\text{D}$ is close to the number of protons which was found to bind to $h$-BN/Rh(111) in cyclic voltammograms in an electrolyte \cite{wid07}.

\begin{figure*}
\includegraphics[width=\textwidth]{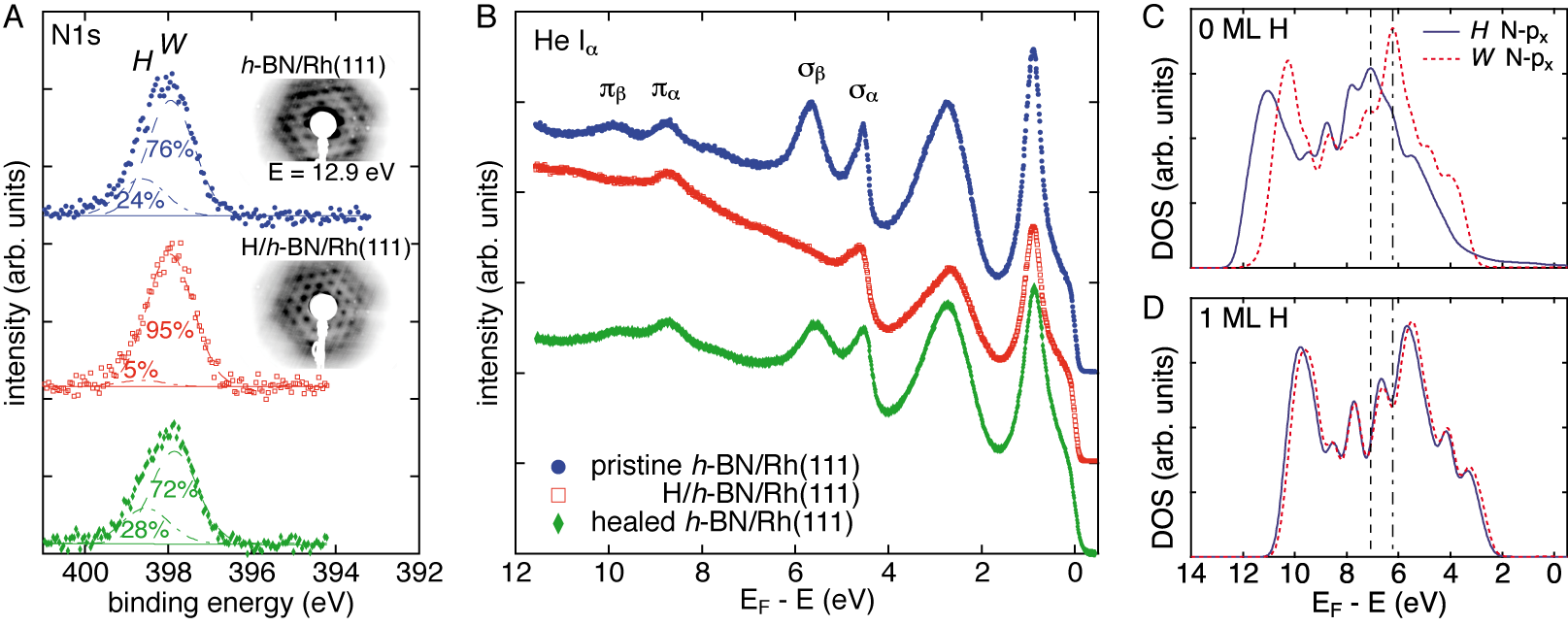}
\caption{\label{F:HhBNRh_PES}A) and B) The $h$-BN/Rh(111) nanomesh before (filled blue circles) and after (open red rectangles) hydrogenation; and after H desorption (filled green diamonds). A) N 1s core level XPS (monochromatized Al K$_\alpha$ radiation) which features a distinct narrowing upon hydrogenation. Insets in A) show the LEED diffraction patterns of the corrugated superstructure of $h$-BN/Rh(111) (upper panel) and the superstructure of H/$h$-BN/Rh(111) (lower panel) for a primary electron energy of $E=\unit{12.9}{\electronvolt}$. B) He I$_\alpha$ normal emission UPS shows a vanishing of the $\sigma_\beta$ and $\pi_\beta$ peaks which are attributed to the holes ($H$) of the $h$-BN/Rh(111) after exposure to H while the $\sigma_\alpha$ and $\pi_\alpha$ peaks corresponding to the wires ($W$) of the $h$-BN/Rh(111) remain.\newline
C) and D) Calculated projected DOS of p$_x$ orbitals originating from N atoms in hole ($H$) and wire ($W$) regions of pristine $h$-BN/Rh(111) and $h$-BN/Rh(111) intercalated with $\unit{1}{ML}$ of H, respectively.}
\end{figure*}
The surface texture switching is also monitored by its spectroscopic signatures that is the \mbox{N 1s} and B 1s XPS spectra and the band splittings of the sp$^2$-derived bands (Fig. \ref{F:HhBNRh_PES}). The evolution of the N 1s core level peak during a switch off -- switch on cycle substantiates the picture (Fig. \ref{F:HhBNRh_PES} A). In the as prepared state the N 1s peak of the nanomesh can be fitted by two Gaussians with a splitting of $\unit{680}{\milli\electronvolt}$ between the low and the high binding energy component corresponding to the wire (N$_W$ 1s) and the hole (N$_H$ 1s) regions, respectively \cite{pre07b}. The fit results in a ratio of about $3:1$ for the peak heights of N$_W$ 1s and N$_H$ 1s. After switching off the surface texture of the nanomesh the peak considerably narrowed at the binding energy of the wire regions. Upon annealing the N 1s peak broadens back to the original ratio of N$_W$ 1s : N$_H$ 1s $\approx3:1$. The same effect is observed for the B 1s core levels (not shown), though the splitting is smaller. This is an indication that the hydrogen does not bind to the $h$-BN, and it is thus conjectured that the hydrogen intercalates, i.e. binds to the Rh substrate. The insets in Figure \ref{F:HhBNRh_PES} A) show low-energy electron diffraction (LEED) data of the nanomesh in the switched on and the switched off state: The super cell size does not depend on the corrugation. The sharpness of the superstructure diffraction spots in the switched off state which originate from the ($13\times13$) $h$-BN on ($12\times12$) Rh units prove the structural integrity of the $h$-BN layer and exclude a twist with respect to the Rh(111) substrate in the flat state. In Figure \ref{F:HhBNRh_PES} B) the He I$_\alpha$ photoemission spectra also indicate a flattening of the $h$-BN layer: The bands related to the holes of the nanomesh ($\sigma_\beta$ and $\pi_\beta$) vanish after H exposure. Furthermore, annealing of the H exposed sample [H/$h$-BN/Rh(111)] to about $\unit{600}{\kelvin}$ recovers the original band splitting as a result of a switch-back to the pristine nanomesh.

In Figure \ref{F:HhBNRh_PES} C) and D) calculated densities of states (DOS) of p$_x$ states on the nitrogen atoms that constitute the $h$-BN $\sigma$ bands are displayed. The results for the clean nanomesh in Figure \ref{F:HhBNRh_PES} C) are in line with the experiment and Refs. \cite{las07,ber07}. The N p$_x$  DOS in the hole and wire sites have pronounced peaks at about $\unit{7}{\electronvolt}$ and $\unit{6}{\electronvolt}$ below $E_\text{F}$, respectively, which reproduce the experimentally observed $\sigma$ band splitting of about $\unit{1}{\electronvolt}$ well. The calculated densities in Figure \ref{F:HhBNRh_PES} D) comprise one monolayer of intercalated hydrogen and feature the experimentally observed disappearance of the  hole derived $\sigma_\beta$ band in $h$-BN/Rh(111) under the exposure to H (or D).
\begin{figure*}
\includegraphics[width=\textwidth]{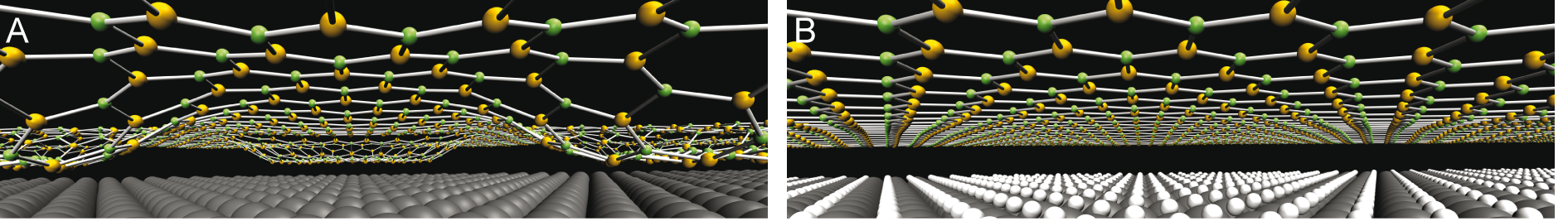}
\caption{\label{F:HhBNRh_DFT}Three dimensional representation of the results from the structure optimizations with DFT. A) The clean $h$-BN/Rh(111). B) The $h$-BN/Rh(111) with one H atom per Rh unit intercalated between the $h$-BN and the topmost Rh(111) layer. In the pictures the corrugation amplitude of the $h$-BN layer has been enhanced by a factor of two. Rh atoms are displayed in grey, B and N atoms in orange and green, respectively, and H atoms in white. Atom sizes are not to scale.}
\end{figure*}
The structure of the full $h$-BN/Rh(111) system (567 Rh atoms plus 169 BN pairs in the unit cell) was optimized by density functional theory (DFT). The results agree with previously reported findings \cite{las07}. The optimized structure is displayed in Figure \ref{F:HhBNRh_DFT} A) and shows a corrugated $h$-BN layer where about $\unit{40}{\%}$ of the BN units are in close and about $\unit{60}{\%}$ are in loose contact to the Rh(111) substrate forming the holes and the wires, respectively [about $\unit{2.2}{\angstrom}$ and $\unit{3.2}{\angstrom}$ above the topmost Rh(111) layer]. 
Hydrogen intercalation below $h$-BN on Rh(111) has been investigated in several steps starting with a single H atom on top of a free standing $h$-BN layer where no binding was found for distances greater than $\unit{2}{\angstrom}$. On the pristine Rh(111) surface H bonding is observed up to at least one monolayer where H atoms favor \emph{fcc} hollow sites in accordance with \cite{fuk07}. Also, distributing atomic H between the $h$-BN and the top Rh layer and optimizing the structure leads to a H--Rh bonding for the case of the full monolayer, H atoms again occupy \emph{fcc} hollow sites. In the final step the full system of $13\times 13$ BN units and a four layer $12\times12$ Rh(111) slab in the presence of intercalated H has been optimized. Figure \ref{F:HhBNRh_DFT} B) shows the resulting structure for $12\times 12$ intercalated H atoms per supercell unit ($\unit{1}{ML}$) which again favor \emph{fcc} hollow sites and level at a height of $\unit{1.0}{\angstrom}$ above the topmost Rh layer. A fundamental feature of the H intercalated system is the drastic reduction of the $h$-BN corrugation amplitude by a factor of five from $\unit{1.1}{\angstrom}$ to $\unit{0.2}{\angstrom}$ which supports the picture of H intercalation drawn from the experiment. A gradual repulsion of the $h$-BN layer was found for increasing amounts of H, where the mean $h$-BN layer -- Rh distance increases from about $\unit{2.9}{\angstrom}$ for the pristine $h$-BN/Rh(111) to about $\unit{3.4}{\angstrom}$ for $h$-BN/Rh(111) intercalated with $\unit{1}{ML}$ of H. This structural change directly influences the electrostatic potential whose corrugation amplitude within the nanomesh unit cell drops from $\unit{480}{\milli\electronvolt}$ to $\unit{180}{\milli\electronvolt}$ at a distance of $\unit{3.9}{\angstrom}$ above the $h$-BN layer. Thus the strength of the lateral dipole rings in the $h$-BN layer, which are located at the sites of the highest corrugation gradient and which are the source of the molecular traps of the $h$-BN/Rh(111) nanomesh, is expected to be reduced by about $90\,\%$ \cite{dil08}. The experimentally observed disappearance of the $\sigma_\beta$ band in UPS after H intercalation is also confirmed by theory where the $\sigma$ band splitting between the hole and wire derived N p$_x$ DOS decreases with the amount of intercalated H, finally rendering both DOS experimentally indistinguishable for coverages $\gtrsim \unit{0.5}{ML}$ (Fig. \ref{F:HhBNRh_PES} D).

\section{Conclusions}

In conclusion it is shown that exposure of the a single layer of hexagonal boron nitride on Rh(111) to atomic hydrogen leads to a change of the surface texture where the corrugation of the sp$^2$ layer vanishes. This flattening is accompanied by a strong change of the electronic structure of $h$-BN/Rh(111). In the flat state at least one H atom is captured between four $h$-BN and Rh(111) units. Mild annealing recovers the original surface texture of the $h$-BN/Rh(111) nanomesh. Regarding similar sp$^2$-hybridized template systems the metallic analog of the nanomesh -- graphene on Ru(0001) -- is expected to show comparable switching behavior.

\bibliography{../../../../references.bib}

\end{document}